# Vanadyl dithiolate single molecule transistors: the next spintronic frontier?

S. Cardona-Serra,[a,b] A. Gaita-Ariño[a]

The role of Chemistry in the road towards quantum devices is the design of elementary pieces with a built-in function. A brilliant example is the use of molecular transistors as nuclear spin detectors, which, up to now, has been implemented only on [TbPc2]-. We argue that this is an artificial constraint and critically discuss the limitations of current theoretical approaches to assess the potential of molecules for their use in spintronics. In connection with this, we review the recent progress in the preparation of highly coherent spin qubits based on vanadium dithiolate complexes and argue that the use of vanadyl dithiolates as single molecule transistors to read and control a triple nuclear spin qubit could give rise to new phenomena, notably including a low-current nuclear spin detection scheme by means of a spin valve effect.

In the new landscape of quantum technologies, rational molecular design is finding its role: the preparation of smart elementary pieces with a minimal but useful built-in function. [1]. Recent progress in this direction has been remarkable. The first ingredient, namely long quantum coherence times, has been markedly improved for molecular electronic spin qubits. Record times are now in the scale of decades or hundreds of microseconds. [2,3] In part this progress has been based on general mechanisms which are already well understood, [4] whereas other tricks such as the use of the so-called Atomic Clock Transitions have only recently been applied in this context;[5] this might still benefit from vibration-focused molecular optimization.[6, 7] In particular, rigid polyoxometalate complexes are being considered as test subjects for simple experiments in Single-Molecule Spintronics and Molecular Quantum Computing.[5, 8, 9] The next step is scaling up and wiring the elementary pieces into complex circuits. Strategies for supramolecular organizations have been arising,[10] as have modular designs of molecular qubits to implement universal quantum gates.[11, 12] Wiring up the molecular pieces is the most challenging step, although some proposals have been made in this direction.[13]

Herein we will discuss two seemingly distant projects of rational design of molecular spin qubits, namely a series of experiments in molecular spintronics for quantum computing and a chemical family of molecules that are being studied as qubits, and argue that it would be beneficial to combine them. What makes these projects special is the fact that each is on the cutting edge of their respective fields. The molecular spintronics setup, which allows reading and manipulating a single nuclear spin, is perhaps the scheme within molecule-based quantum computing that is closest to achieving a minimal functional quantum algorithm. In parallel, $V(C_8S_8)_3^{2-}$, a vanadium(IV) dithiolate derivative, is the molecular complex with the longest electron-spin decoherence time ($T_2$ = 0.7 ms in optimized conditions).

The molecular spintronic setup we are referring to is the use of molecular transistors as nuclear spin detectors, which has been experimentally achieved using the 4 nuclear spin states of the bis-phthalocyaninato $Tb^{III}$ complex $[Tb(Pc)_2]^-$ [14, 15]. This approach, developed by the group of prof. Wernsdorfer, consists in observing a jump in the spin dependent conductance when the system passes through an electronuclear anticrossing at a magnetic field determined by its nuclear spin projection. Within this experimental study of a single-molecule device where the nuclear spin and its hyperfine coupling are related with an external spin current, major achievements include: (a) the manipulation of such hyperfine interaction by the external application of an electric field[16] (b) the combination of such effect with a certain crystal field environment, in order to suppress quantum tunneling of magnetization at zero field in single ion magnets (allowing the protection of quantum information in that qubit) [17], (c) the possibility to scale the approach into a two qubit gate which can be rationally organized in a surface, [18] (d) the study of nuclear spin isomers and relationship between the magnetic relaxation and other nuclear-spin-driven events. [19] and, (e) the combination of this electronic read-out with the application of a transversal magnetic field which has permitted the quantitatively evaluation of the interaction between the electronic and the nuclear spin. [20] Overall, this proposal opens the door for the use of nuclear and electron spin resonance techniques to perform basic quantum operations.

Almost all experiments in this area were carried out with the same molecular complex, $[Tb(Pc)_2]^-$, and it is important to understand that this was mostly out of convenience and efficiency. A myriad of further systems could be just as useful, as long, as, first, anticrossings between electronuclear spin states exist and are accessible, and second, even in the transport regime the molecular spin is localized. There are a few other rather obvious desiderata in terms of chemical stability and affinity, as discussed elsewhere. [21] The problem is the theoretical difficulty in making reliable predictions, and the challenging experimental setup, which means once a molecule works, the most convenient path is to keep using the same system.

One needs to admit that theoretical calculations concerning spin transport and magnetic scattering in molecules between electrodes are extremely challenging due to their complexity. In a first approximation, the Landauer approach seemed to be sufficient to explain the transmission pathways followed by the electronic current, e.g. this permitted to explain the behavior of a spin crossover dimer under the effect of the bias voltage. [22, 23] But this is in fact insufficient for other spintronic consequences, as was shown in an ingenious contribution by Lorente et al. [24] which was instrumental in unveiling the most common failures in the assignment of the spin filter behavior, specially those regarding the broken symmetry description of the spin state. Additionally, the HOMO / LUMO levels can be pinned to the Fermi energy thus giving a erroneous view of the transmission levels. This problem has been overcome by the use of the 'atomic self-interaction correction' which shows a better agreement with benchmark experiments.[25] Even today, there is a lack of theoretical tools that are able to predict the molecular conduction properties for each specific single molecule device. Within these limitations, in a series of recent theoretical works we performed theoretical simulations of single magnetic molecules located between two gold electrodes. [1, 21, 26] Thus we generalize experimental results by the group of prof. Wernsdorfer, involving molecular nuclear-spin-transistor (see Fig 1, left), defining the chemical requirements for this behavior. These calculations are based on standard tools and could be applied to very many potential systems. In our first example, we studied a set highly coherent of vanadium dithiolate complexes. (see i.e. Fig 1, right)

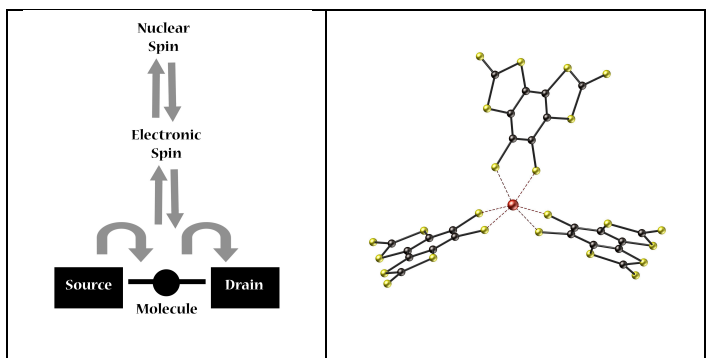

Figure 1: Left: simplified scheme for a nuclear-spin sensitive molecular transistor between two gold electrodes. Right: high-$T_2$ molecular spin qubits, which is also endowed with terminal sulfur atoms.

This brings us to the second project: the use of dithiolate-based mononuclear complexes in the rational design of coherent qubits. These molecular entities are inherently promising for that goal: dithiolate complexes present few or no nuclear spins in the vicinity of the electron spin, thus minimizing magnetic noise which is detrimental to quantum coherence, and they typically bind to the metal acting as rigid, aromatic 'blades', diminishing the spin-vibrational coupling,[27] which further protects the spin from coherence. This allowed the groups of van Slageren and Freedman to establish subsequent records in highly coherent qubits, and indeed these systems still are the most coherent among magnetic molecules. [2, 3, 28] An excellent report on recent progress of mononuclear transition metal complexes as spin qubits, highlighting the work of these two groups, was provided recently by Sproules. [29] This line of work has been fertile: in the rich hyperfine Hilbert space of vanadium dithiolate complexes long coherence times that persist at high temperature have been observed, $T_2$ = 1.2 μs at 80 K, as well as quantum coherences from multiple transitions. [30] From the point of view of the chemistry, one can prepare small and simple dithiolate complexes and this facilitates crystal engineering by playing with the counteraction, as illustrated in the case of $[Co(C_3S_5)_2]^{2-}$, where the countercation-regulated dihedral angle between the two ligand 'blades' was used to adjust the Hamiltonian and thus vary the spin dynamics, from single-molecule magnet behavior to qubit behavior. [31] At the same time, dithiolate chemistry provides enough flexibility to allow for systematic studies by altering the ligand, for example preparing ligands with essentially the same planar structure and carbon-sulfur based but increasing 'blade' length. [32, 33] Considering them from a broader perspective, these systems present a rich chemistry which extends well beyond the few studies in the field of molecular spin qubits.[34, 35, 36] It is clear that they offer enough tunability to allow sufficient tailoring of their chemical and spintronic behavior. In particular, these systems are known to attach strongly to a gold surface due to the well-characterized sulfur-gold bond, [37] and the spin has been to survive the loss of electrons [38]

Let us now discuss the pros and cons of dithiolate complexes as molecular component in a molecular spin transistor. Here we will consider vanadium complexes, not only because a vanadium dithiolate complex currently holds the record for quantum coherence but also because vanadium has 8 nuclear spin states, and a rich nuclear spin structure of the magnetic ion involves the possibility to encode a larger number of qubits: $2^n$ levels translate into n qubits. It is then critical to note that dithiolate derivatives complexing vanadium ions can be naïvely divided into two sets. The first contains the complexes in which the $V^{IV}$ is coordinated by three dithiolate derivatives. In the other set, two ligand molecules coordinate the vanadium ion, which completes its coordination with an oxygen atom, resulting in a VO vanadyl moiety; the stability of the vanadyl moiety is then high enough to stop further substitution. In terms of single molecule experiments, this latter group seems particularly interesting because of the fewer possibilities of attachment between two electrodes.

This is not the end of the story. There is a vital parameter for the design of the ideal vanadium complex, namely the hyperfine coupling, which, as we will see, determines the playing field for sensing and manipulating the nuclear spin qubit. One needs to emphasize that, in this context, different kinds of spintronic experiments have different requirements on the molecule. For the detection of the nuclear spin, widely equispaced anticrossings at identical frequencies translate into non-overlapping, characteristic magnetic fields for different anticrossings, which facilitates detection. On the contrary, in absence of the characteristically large quadrupolar coupling of TbPc2, clearly distinguishable transition frequencies between different nuclear spin levels are necessary for the nuclear spin state manipulation at a given magnetic field. Sadly, these two conditions are in contradiction with each other. In both cases, the key lies in the hyperfine interaction, and in particular in the perpendicular to parallel ratio $H_{PER}/H_{PAR}$. For clarity, four qualitative schemes for representative ratios are depicted in Figure 2.

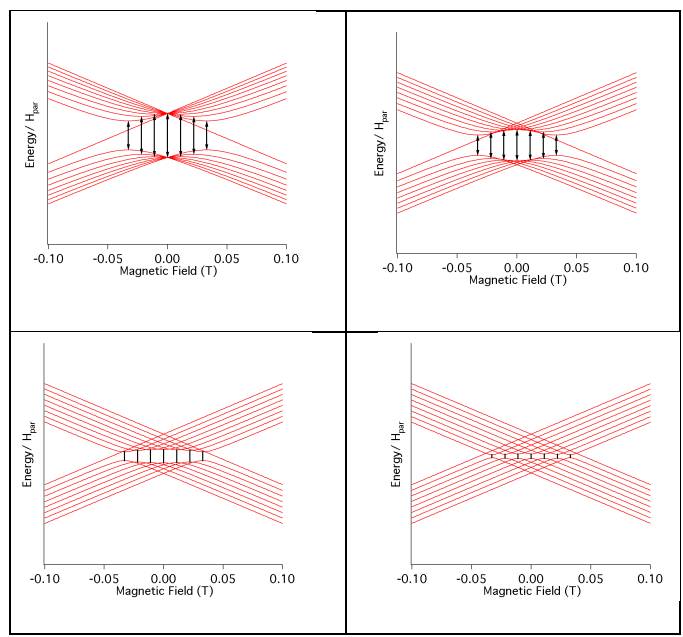

Figure 2: Evolution of the hyperfine energy level scheme for idealized vanadium complexes, from closely bunched anticrossings at very different frequencies to widely equispaced anticrossings at identical frequencies. Parameters: (top, left) $H_{PER}/H_{PAR}=1$, (top, right) $H_{PER}/H_{PAR}=0.7$, (down, left) $H_{PER}/H_{PAR}=0.3$, (down, right) $H_{PER}/H_{PAR}=0.1$. Black lines represent the avoided crossings between the electronuclear levels.

To illustrate the effect of hyperfine coupling, the ratio $H_{PER}/H_{PAR}$ is increased from 1/10 to 1/1. This produces a continuous evolution from closely bunched anticrossings at very different frequencies to widely equispaced anticrossings at identical frequencies, with intermediate schemes at intermediate ratios. Typical values for vanadyl present a ratio of approximately 1/4 [39] whereas typical values for vanadium complexes move in ratios around 5/1 [3]. With the conditions described above, the ideal complex would be a vanadyl where, by chemical design, perhaps in combination with an external electrical field, an $H_{PER}/H_{PAR} \approx 0.7$ ratio has been achieved, an ideal compromise between nuclear spin addressing and nuclear spin reading. This is by no means an obvious task, since the very deposition between electrodes, including the presence or absence of counterions, might influence this ratio. In the absence of a specific chemical design, vanadyl complexes, with anticrossings that appear at clearly differentiated fields, will tend to excel for spin detection and vanadium complexes, with transitions presenting clearly different frequencies, will be ideal for spin manipulation.

As we want to open a discussion on the potential synergies between these spintronic experiments and vanadium dithiolates molecules, let us start by describing one of the challenges in this setup, namely Joule heating: a current is circulating through a physical system which is otherwise at extremely low temperatures. In the case of nuclear spin sensing via molecular transistors, the temperatures are typically kept at temperatures that range between 40 mK and 150 mK.[14] Even with modest voltages, every passage of a single electron through a voltage difference of 1 mV results, via Joule heating, in an energy of 1 meV in the molecular junction, which needs to be dissipated. And even with modest currents such as usually employed in these experiments, 1 nA means $6\cdot10^9$ electrons/second. Indeed, it is known that nuclear spin temperature increases monotonically with 'drain-source' voltage, with the increase in the nuclear spin temperature being attributed to energy exchange with the electrons passing through the molecule.[15] Even extremely low currents involve the dissipation of non-negligible energies from the very small volume occupied by the molecule plus the two leads. This can lead to difficulties in the stability of the temperature and the potential for atomic-scale rearrangements -either in the leads, or in the molecule- which effectively change the experimental parameters. In fact, the current flow through the molecule is only desirable at the moment of the electronic spin flip, where the change in conductance evidences this transition and permits the determination of the nuclear spin state. It would therefore be ideal to have a scheme where the current flow is limited precisely to that point. In principle, this could be achieved via a spin valve.

To fulfill these requirements, it should be possible to chemically design a vanadyl dithiolate complex which acts a double spin filter and is able to conduct spin up (down) carriers depending on a fine tuning of the gate voltage. Indeed, as a mononuclear magnetic complex, this molecule presents a well-defined density of states;[21] moreover their chemical tunability allows arranging transmission channels to function as a double spin filter.[26] Thus, we can imagine a single molecule experiment where the molecular wire admits the transmission of spin up carriers. When the sweeping magnetic field passes through the electronuclear anticrossing and flips the magnetic momentum, the system becomes momentarily conductive. Then, an automated change on the gate voltage would return the system to a high resistance state by only allowing the opposite spin carriers. The back and forth sweeping of the magnetic field would then allow to measure the nuclear spin in the sample without heating it up with a continuous flow of electric current. (See Figure 3).

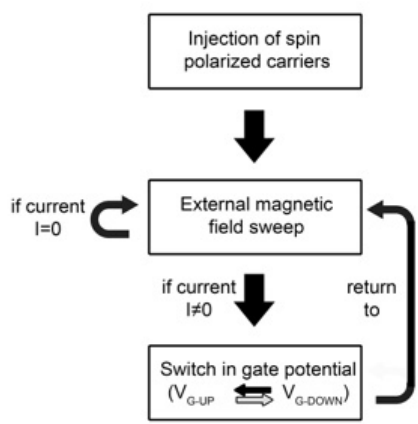

Figure 3: A trivial application of the clear transmission spectra which are characteristic of mononuclear complexes: using a dual spin filter for nuclear-spin sensitive molecular transistor with minimal current.

## Conclusions

It seems clear that there is no fundamental reason to limit nuclear spin molecular transistors, arguably the most cutting edge experiment in the molecular spin qubit field, to the currently employed Single Ion Magnet. Vanadyl dithiolates are a promising and rich family in molecular magnetism, which hold the current record of coherence and seem promising in this new context. [40] If they can be successfully contacted in break junction setups, one can envision that new kinds of experiments would be made possible, of which we gave here an explicit example, namely, the read-out of the nuclear spin state by using a single-molecule transistor but with minimal current flow.

## Conflicts of interest

There are no conflicts to declare.

## Acknowledgements

The research reported here was supported by the Spanish MINECO (Grants MAT 2014-56143-R and CTQ 2014-52758-P co-financed by FEDER, and Excellence Unit María de Maeztu MDM-2015-0538), the European Union (ERC-CoG DECRESIM 647301 and COST-MOLSPIN-CA15128 Molecular Spintronics Project), and the Generalitat Valenciana (Prometeo Program of Excellence). A. Gaita-Ariño thanks the Spanish MINECO for a Ramón y Cajal Fellowship.

## Notes and references